\newcommand{\submissionver}{1}
\newcommand{\journalsubmission}{0}

\ifnum\submissionver=1%
    \ifnum\journalsubmission=1%
        \documentclass[AER,reviewmode]{AEA}%
    \else%
        \documentclass[AER,draftmode]{AEA}%
    \fi
\else%
    \documentclass[AER,draftmode]{AEA}%
\fi%

\usepackage{natbib}

\usepackage{graphicx}
\usepackage{quoting}

\newtheorem{theorem}{Theorem}
\newtheorem{proposition}{Proposition}
\newtheorem{corollary}{Corollary}
\newtheorem{remark}{Remark}
\newtheorem{assumption}{Assumption}
\newtheorem{example}{Example}

\draftSpacing{1.5}

\begin{document}

\ifnum\submissionver=1%
    \title{Manipulation-Robust Regression Discontinuity Designs}%
    \shortTitle{Manipulation-Robust Regression Discontinuity Designs}%
\else%
    \title{}%
    \shortTitle{}%
\fi%

\author{Takuya Ishihara and Masayuki Sawada\thanks{Ishihara: Graduate School of Economics and Management,
  Tohoku University, 27-1 Kawauchi, Aoba-ku, Sendai, Miyagi 980-8576, Japan, takuya.ishihara.b7@tohoku.ac.jp. 
Sawada: Institute of Economic Research,
  Hitotsubashi University, 2-1 Naka, Kunitachi,
Tokyo 186-8601, Japan, m-sawada@ier.hit-u.ac.jp. We would like to thank Michihito Ando, Yoichi Arai, Yu Awaya, Marinho Bertanha, Kojima Fuhito, Kentaro Fukumoto, Koki Fusejima, Hidehiko Ichimura, Masaaki Imaizumi, Guido Imbens, Shoya Ishimaru, Ryo Kambayashi, Kohei Kawaguchi, Nobuyoshi Kikuchi, Toru Kitagawa, Kenichi Nagasawa, Shunya Noda, Ryo Okui, Tomasz Olma, Shosei Sakaguchi, Pedro Sant'Anna, Yuya Sasaki, Katsumi Shimotsu, Suk Joon Son, Kensuke Teshima, Edward Vytlacil, Junichi Yamasaki, Takahide Yanagi, and the participants of the SWET 2020, Happy Hour Seminar!, Kansai Keiryo Keizaigaku Kenkyukai 2021, ASEM 2021, Ouyou Keiryoukeizaigaku Conference 2021, IAAE 2022 Annual Conference, ESAM 2022, seminars at Kyoto University, Hitotsubashi University, and the University of Tokyo for their valuable comments. This study was supported by a Grant-in-Aid for JSPS KAKENHI (Grant Number 22K13373 (Ishihara) and 21K13269 (Sawada)).}} 
\date{\today}
\pubMonth{}
\pubYear{}
\pubVolume{}
\pubIssue{}
\JEL{}
\Keywords{Regression Discontinuity Design; Manipulation; Diagnostic Test}

\begin{abstract}
We present simple low-level conditions for identification in regression discontinuity designs using a potential outcome framework for the manipulation of the running variable. Using this framework, we replace the existing identification statement with two restrictions on manipulation. Our framework highlights the critical role of the continuous density of the running variable in identification. In particular, we establish the low-level auxiliary assumption of the diagnostic density test under which the design may detect manipulation against identification and hence is manipulation-robust.
\end{abstract}

\maketitle

\ifnum\submissionver=1%
\else%
\newpage 
\fi%

The credibility of regression discontinuity (RD) design, ``one of the most credible non-experimental strategies for the analysis of causal effects'' (\citealp{cattaneo_idrobo_titiunik_2020}), is questioned when individuals \textit{manipulate} the running variable that determines the treatment assignment. Hence, assessing manipulation is a critical procedure in RD analyses. However, the definition for manipulation is absent in existing models. 

As manipulation has not been defined, identification argument is silent about manipulation. Thus, one cannot effectively argue for identification against a particular manipulation in consideration. In other words, existing identification is based on high-level conditions.

In this study, we articulate identification under low-level conditions so that it can be explicitly justified under manipulation. We establish the conditions using a potential outcome framework for the running variable that determines treatment assignment. Specifically, we define manipulation as an indicator that assigns one of two potential running variables, with and without manipulation. Given this framework, we restate the \textit{continuity} condition \citep{hahnIdentificationEstimationTreatment2001} for identification with simple and explicit low-level conditions.

For example, a test score is a running variable when passing the examination is treatment. Manipulation is an action that assigns a manipulated score for the manipulated student and a non-manipulated score for other students. With manipulation, only one of two scores is observed. For instance, a teacher manipulates a test score of a student who has a non-manipulated failing score to a manipulated passing score. For such a manipulated student, we observe only the manipulated score not their original non-manipulated score. 

Our low-level conditions require that manipulation must be a randomization. Specifically, manipulation must not only randomly assign the manipulated score when it is around the passing cutoff but also randomly select students to manipulate when their original non-manipulated scores are around the passing cutoff. In other words, the teacher must not \textit{precisely assign} the score to pass the examination nor \textit{precisely select} students who have failing scores. 

Formally, we introduce an indicator of manipulation, $M$, such that the realized running variable is given by $R = M R^*(1) + (1 - M) R^*(0)$, where $R^*(1)$ is the running variable with manipulation ($M=1$) and $R^*(0)$ is the running variable without it ($M=0$). Identification holds under two restrictions: The manipulation $M$ randomly assigns $R = R^*(1)$ when $R^*(1)$ is assigned around the cutoff and randomly selects units to manipulate when they have $R^*(0)$ around the cutoff. In other words, manipulation must not \textit{precisely} assign $R^*(1)$ around the cutoff nor \textit{precisely} select units whose $R^*(0)$ are around the cutoff. Those two restrictions accommodate most examples, as illustrated in previous studies such as \cite{leeRegressionDiscontinuityDesigns2010} and \cite{gerardBoundsTreatmentEffects2020}. Our framework formalizes their examples as two simple and explicit restrictions on $(M, R^*(0), R^*(1))$.

In this framework, the continuous density function is not a by-stander of identification. We demonstrate that continuous density is critical for the continuity condition \citep{hahnIdentificationEstimationTreatment2001}, which is decomposed into the continuous mean potential outcomes \textit{weighted} by the continuous density functions. Furthermore, we establish a low-level condition for the density test because although ``a running variable with a continuous density is neither necessary nor sufficient for identification except under auxiliary assumptions'' \citep{mccraryManipulationRunningVariable2008}, the auxiliary assumption was absent. 

We provide the auxiliary assumption as two restrictions on \textit{precise} manipulation: If manipulation precisely assigns $R^*(1)$ or precisely selects units who have $R^*(0)$ around the cutoff, it must not assign to and select from both sides of the cutoff. In other words, precise manipulation must be \textit{one-sided} both in its assignment of $R^*(1)$ and selection in $R^*(0)$. This one-sided restriction is also a low-level condition for partial identification in \cite{gerardBoundsTreatmentEffects2020} where a special case of their result justifies the density test. \footnote{Their restriction directly relies on local randomization and hence is also a high-level condition. See Remark \ref{rem.grr} in Section \ref{sec.mrrd} for a detailed discussion.} In summary, our framework induces low-level conditions for identification, diagnostic tests, and partial identification under possible manipulation. In other words, researchers may assess whether their RD designs are \textit{manipulation-robust}.

We emphasize that although the existing practices are valid, their usage must be updated based on new interpretations. Statistical packages for RD design are well established. For example, \textit{rdrobust} (\citealp{calonicoRobustNonparametricConfidence2014a}, \citealp{Calonico_Cattaneo_Farrell_Titiunik_2017}) is the dominant option for estimation and balance or placebo tests, while the \textit{rddensity} package (\citealp{Cattaneo_Jansson_Ma_2018}, \citealp{cattaneoSimpleLocalPolynomial2019}) is increasingly used for the density test. Bounds are available from \textit{rdbounds} package \citep{gerardBoundsTreatmentEffects2020}. Although all these devices remain functional, our framework clarifies when and how they are used under what assumptions.

\section{Related literature and our contributions}
RD is a powerful tool in various disciplines. For extensive surveys, see \cite{leeRegressionDiscontinuityDesigns2010}, \cite{DiNardo_Lee_2011}, \citeauthor{cattaneo_idrobo_titiunik_2020} (\citeyear{cattaneo_idrobo_titiunik_2020},\citeyear{Cattaneo_Idrobo_Titiunik_2024}). Among the growing body of RD studies, we contribute to two strands of the literature. 

First, we contribute to the literature on point identification in RD designs. \citet{hahnIdentificationEstimationTreatment2001} formalize the idea of RD in \cite{Thistlethwaite_Campbell_1960} with the minimal but less intuitive continuity condition.
\cite{leeRandomizedExperimentsNonrandom2008} replaces the continuity condition with an analogy of randomization: If the running variable is equally likely to be just below or above the cutoff then the treatment is as if randomized. \cite{leeRegressionDiscontinuityDesigns2010} illustrate a local randomization failure under manipulation that precisely assigns the running variable to the right of the cutoff. However, manipulation is implicit in their identification argument because it is not defined in their model. \cite{mccraryManipulationRunningVariable2008} introduces an explicit manipulation concept although its connection to identification remains implicit. In this study, we provide the first low-level identification condition with explicit restrictions on manipulation by formalizing the concepts introduced by \cite{mccraryManipulationRunningVariable2008}, \cite{leeRandomizedExperimentsNonrandom2008}, and \cite{leeRegressionDiscontinuityDesigns2010} via a potential outcome framework.

Second, we contribute to the literature on diagnostic tests for continuous density function (density test) and continuous conditional mean functions of covariates (balance or placebo test) as consequences of local randomization \citep{leeRandomizedExperimentsNonrandom2008}. These tests have been updated. \citet{otsuEstimationInferenceDiscontinuity2013} propose an empirical likelihood test for the density test, \citet{cattaneoSimpleLocalPolynomial2019} propose a density test from local polynomial estimates, \cite{bugniTestingContinuityDensity2020a} propose a density test with $g$-order statistics, and \cite{canayApproximatePermutationTests2018} propose randomization tests for covariate balancing. However, the null hypotheses of these tests do not imply identification. \footnote{\cite{mccraryManipulationRunningVariable2008} conjectures an assumption (\textit{monotonic} manipulation) for the density test, but the conjectured condition is neither necessary nor sufficient. See the Supplementary Appendix for details.} For fuzzy designs, \citet{Arai_Hsu_Kitagawa_Mourifie_Wan_2021} propose a complementary diagnostic procedure that is free from restrictions on the density. \footnote{\cite{Bertanha_Imbens_2020} also discuss tests for exogeneity and external validity in fuzzy designs.} However, there is no testable restriction in their procedure for sharp designs and the reduced form of fuzzy designs. \cite{gerardBoundsTreatmentEffects2020}'s condition for partial identification may be used for the density test as a special case. Because their condition is a high-level assumption that can be challenging to justify for some designs, our auxiliary assumption is the first low-level condition to connect these diagnostic tests with identification.

\section{Low-level identification conditions}\label{sec.identification}

In an RD design, we exploit treatment assignment $D \in \{0,1\}$ by a scalar running variable $R$ by exceeding the cutoff $c$. For example, students receive qualification $D$ when their test score $R$ exceeds the passing cutoff $c$. This treatment $D = 1\{R \geq c\}$ represents a sharp design. \footnote{We do not consider measurement error for $R$ that has been studied extensively. For example, see \cite{Yu_2011}, \cite{Davezies_Le_Barbanchon_2017}, \cite{Pei_Shen_2017}, \cite{Yanagi_2017}, \cite{Bartalotti_Brummet_Dieterle_2020} and \cite{Dong_Kolesar_2022}.} For fuzzy designs with noncompliance, we consider their reduced form, essentially a sharp design. For a pair of potential outcomes, $\{Y(1),Y(0)\}$, $Y = DY(1) + (1 - D)Y(0)$ is observed. We aim to identify our target parameter, the average treatment effect (ATE) for students whose score is at the cutoff: $E[Y(1) - Y(0)|R=c]$. 

The remainder of this paper introduces the following notation: For a random variable $Z$, let $E[Z|R=c_+] \equiv \lim_{r \downarrow c} E[Z|R=r]$ and $E[Z|R = c_-] \equiv \lim_{r \uparrow c} E[Z|R=r]$. For a density function $f$, let $f(c_+) \equiv \lim_{r \downarrow c} f(r)$ and $f(c_-) \equiv \lim_{r \uparrow c} f(r)$. Throughout this paper, we impose Assumption \ref{ass.regularnoG} in Appendix, which imposes the existence of relevant moments, densities, and their limits.

Identification of the ATE follows from the \textit{continuity} condition \citep{hahnIdentificationEstimationTreatment2001}
\begin{equation}
 E[Y(d)|R=c_-] = E[Y(d)|R=c_+] \mbox{ for } d \in \{0,1\}.  \label{eq.HTV}
\end{equation}
This condition is the minimal restriction for identification under an ideal design with the same mean types $Y(d)$ for those who have $R$ near the cutoff $c$. A concrete mechanism for \eqref{eq.HTV} is local randomization of $R$ \citep{leeRandomizedExperimentsNonrandom2008}. \footnote{This local randomization differs from a related recent concept in \cite{Cattaneo_Frandsen_Titiunik_2015} and \cite{Cattaneo_Titiunik_Vazquez-Bare_2017} who consider explicit randomization within a small range near the cutoff.} In \cite{leeRandomizedExperimentsNonrandom2008}, if $R$ is \textit{locally randomized}, the following restrictions hold
\begin{equation}
 f_R(c_+) = f_R(c_-) \mbox{, and } E[Y(d)|R=c_-] = E[Y(d)|R=c_+] \mbox{ for } d \in \{0,1\} \label{eq.localrandomized}
\end{equation}
where $f_R(r)$ is the density for $R$. Nevertheless, local randomization is also a high-level condition for an ideal design. These restrictions are silent about manipulation of the running variable because manipulation is not defined in their models.

In this study, we provide simple low-level conditions for \eqref{eq.HTV} from the potential outcome framework for manipulation $(M, R^*(1), R^*(0))$. As in the Introduction, $M \in \{0,1\}$ is an unobserved indicator of manipulation, $R^*(1)$ is the counterfactual running variable with manipulation, and $R^*(0)$ is the counterfactual running variable without manipulation. Hence, $R = M R^*(1) + (1 - M)R^*(0)$. In this framework, identification is shown under the low-level condition where both manipulated $R^*(1)|M=1$ and non-manipulated $R^*(0)|M=0$ are \textit{locally randomized} as in \eqref{eq.localrandomized}. In other words, manipulation must be a random assignment of $R^*(1)$ when it assigns $R^*(1)$ around $c$, and also a random selection of units whose $R^*(0)$ are around $c$.
\begin{proposition} \label{prop.identify}
For each $d \in \{0,1\}$, if
\begin{align}
 &f_{R^*(1)|M=1}(c_-) = f_{R^*(1)|M=1}(c_+), \label{eq.control}\\
 &E[Y(d)|R^*(1)=c_-,M=1] = E[Y(d)|R^*(1)=c_+,M=1] \nonumber\\
 & \mbox{and} \nonumber \\
 &f_{R^*(0)|M=0}(c_-) = f_{R^*(0)|M=0}(c_+), \label{eq.selection}\\
 &E[Y(d)|R^*(0)=c_-,M=0] = E[Y(d)|R^*(0)=c_+,M=0],\nonumber
\end{align} \noindent
then the continuity condition \eqref{eq.HTV} holds and ATE is identified.
\end{proposition}

\begin{proof}
From the potential outcome framework, we achieve the following decomposition:
\begin{equation*}
\begin{split}
 E[Y(d)|R=r] =& E[Y(d)|R^*(1)=r,M=1] P(M=1|R=r)\\
    &+ E[Y(d)|R^*(0)=r,M=0] P(M=0|R=r).
\end{split}
\end{equation*}
Applying Bayes's rule, $P(M=m|R=r) = f_{R|M=m}(r)P(M=m)/f_R(r)$ for $m \in \{0,1\}$, we obtain the following expressions where each component is continuous at $r=c$
\begin{equation}
\begin{split}
 E[Y(d)|R=r] =& E[Y(d)|R^*(1)=r,M=1]\\
 & \hspace{1.0in} \times f_{R^*(1)|M=1}(r) P(M=1)/f_R(r) \label{eq.decomp}\\
    &+ E[Y(d)|R^*(0)=r,M=0] \\
    & \hspace{1.0in} \times f_{R^*(0)|M=0}(r) P(M=0)/f_R(r)
\end{split}
\end{equation}
because
\[
 f_R(r) = f_{R|M=1}(r)P(M=1) + f_{R|M=0}(r)P(M=0)
\]
is continuous at $r = c$. Hence, $E[Y(d)|R=r]$ is continuous at $r=c$.
\end{proof}

The decomposition \eqref{eq.decomp} has a critical implication in the relationship between continuous density functions and identification: Identification relies on the balanced mean types $E[Y(d)|R=c_-,M=m] = E[Y(d)|R=c_+,M=m]$ \textit{weighted} by balanced densities $f_{R|M=m}(c_-) = f_{R|M=m}(c_+)$ for each manipulation status $m \in \{0,1\}$. Conversely, identification can fail because of systematic differences in the (non-)manipulator's mean types $E[Y(d)|M=m,R=r]$ or their densities $f_{R|M=m}(r)$ around the cutoff. We illustrate an implication for the latter in an example of a test score $R$ as the running variable for a qualification awarded via the examination as treatment.
\begin{example}\label{ex.almostnec2}
\cite{jepsenLaborMarketReturns2016} document discontinuous density from an opportunity to retake the examination. Examination retakes are manipulations that replace the initial test score $R^*(0)$ with a retaken score $R^*(1)$ such that we observe $R = M R^*(1) + (1 - M)R^*(0)$. The discontinuity arises because only those who initially failed $R^*(0) < c$ retake the examination. Suppose the initial score $R^*(0)$ is randomized such that $E[Y(0)|R^*(0)=c_-] = E[Y(0)|R^*(0)=c_+]$. An obvious failure in \eqref{eq.selection} is a self-selection story: ``Those who choose to retake the test would be expected to differ from those who do not'' \citep{jepsenLaborMarketReturns2016}. Thus, if retaking students $E[Y(d)|R^*(0)=c_-, M=1]$ differ from non-retaking students $E[Y(d)|R^*(0)=c_-, M=0]$, \eqref{eq.selection} can fail because the self-selected $E[Y(d)|R^*(0)=c_-,M=0]$ differ from its original population $E[Y(d)|R^*(0)=c_-]$ which is the same as $E[Y(d)|R^*(0)=c_+] = E[Y(d)|R^*(0)=c_+,M=0]$. Conversely, it may sound innocuous if their mean outcomes are the same. For example, if examination retakes are randomly offered to the subset of those who initially failed, we have $E[Y(d)|R^*(0)=c_-, M=1] = E[Y(d)|R^*(0)=c_-, M=0]$. However, identification still fails because $f_{R^*(0)|M=0}(c_-) < f_{R^*(0)|M=0}(c_+) = f_{R^*(0)}(c_+)$ and $E[Y(d)|R^*(1)=r,M=1]$ and $E[Y(d)|R^*(0)=r,M=0]$ are \textit{weighted} differently around the cutoff. Namely, \textit{those who happen to be around the cutoff by retaking the test, $\{R^*(1) = c_-, M = 1\}$, would be expected to differ from those who were initially around the cutoff, $\{R^*(0)=c_-\}$}. Here, $E[Y(d)|R^*(1)=c_-, M=1]$ is likely to differ from $E[Y(d)|R^*(0)=c_-, M = 0]$ because the former students $\{R^*(1)=c_-,M=1\}$ can have any $R^*(0) < c$ but the latter students must have $R^*(0) = c_-$; hence, the mean outcomes should differ as lower score $R^*(0)$ may be correlated with lower outcome $Y(d)$. Thus, continuous density is critical for identification.
\end{example}

\section{Manipulation-Robust RD designs}\label{sec.mrrd}
From Proposition \ref{prop.identify}, continuous density functions are critical for identification. Continuous density is neither necessary nor sufficient for identification in general. Sufficiency is particularly important for the density test \citep{mccraryManipulationRunningVariable2008} because it guarantees that the null of the density test implies identification. 

In the following, we provide the conditions for the sufficiency, specifically, those under which manipulation is \textit{detectable} and the design is \textit{manipulation-robust} because we may detect if it exists.

We illustrate how detection works and fails using the following toy example of a bribed teacher, inspired by \cite{diamond2016long}. The teacher receives a rebate of $\beta_i > 0$ if a student $i$ with $R_i^*(0)$ passes. Simultaneously, manipulating the score incurs a marginal cost of $\kappa > 0$ where $R_i^*(0)$ is ex-ante randomized locally at the cutoff. In this toy model, the teacher maximizes the net payoff $\beta_i 1\{R_i \geq c\} - \kappa |R_i - R^*_i(0)|$ for each student. For a student $i$ with the initial score $R_i^*(0)$, the teacher's optimal manipulation $M_i$ satisfies
\[
 M_i = 1\{R^*_i(0) < c \mbox{ and } \beta_i \geq \kappa \cdot (c - R^*_i(0))\}
\]
and the optimal $R^{*}_i(1)$ equals $c$. The induced potential outcome framework for manipulation ($M_i,R^*_i(1),R^*_i(0)$) violates both conditions \eqref{eq.control} and \eqref{eq.selection}. The condition \eqref{eq.selection} fails because the teacher precisely selects those who failed $R^*_i(0) < c$ only if they are worth saving $\beta_i \geq \kappa \cdot (c - R^*_i(0))$. The condition \eqref{eq.control} fails because the teacher precisely assigns $R_i^*(1)$ just at the threshold $c$. Hence, neither $R_i^*(0)|M=0$ nor $R_i^*(1)|M=1$ is randomized locally at the cutoff.

\begin{figure}
    \centering
    \ifnum\submissionver=1%
        \includegraphics[width=\columnwidth]{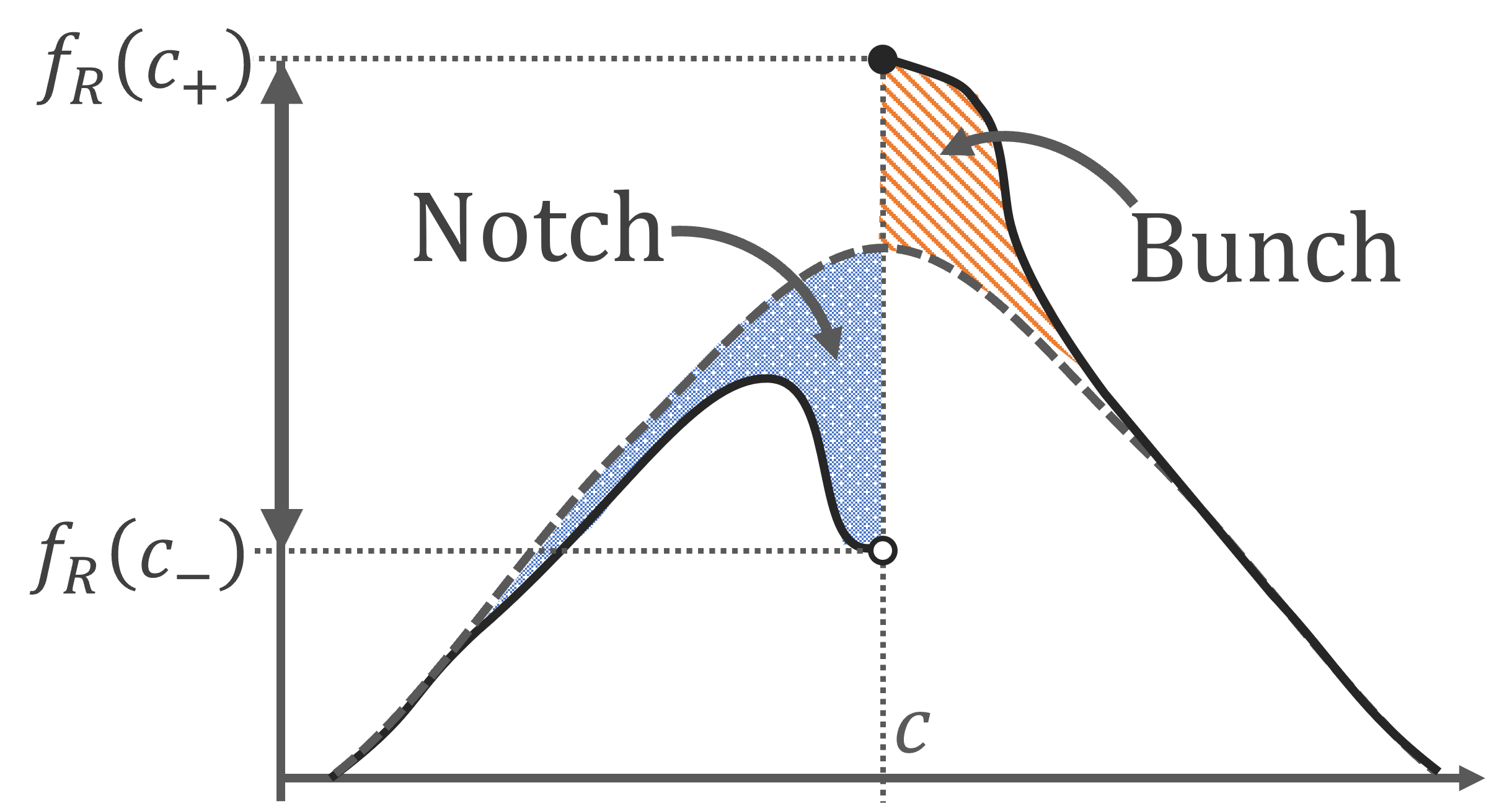}%
    \fi
    \caption{Illustration of a detection idea.}
    \label{fig:illust}
    \begin{figurenotes}
        The dashed line represents the density of $R^*(0)$; the solid line represents the density of $R$, $f_{R}(\cdot)$. From manipulation, the observed density is discontinuous at $c$ due to a bunch (shaded area) into $R=c_+$ and a notch (dotted area) from $R^*(0)=c_-$.
    \end{figurenotes}
\end{figure}

Such manipulation is detectable. Figure \ref{fig:illust} illustrates the initial score $R^*(0)$ and the realized score $R$, deviating from the model by modifying $R^{*}(1)$ so that $R$ has a density. The dashed line represents the density of $R^*(0)$ and the solid line represents the density of $R$. These densities differ in two ways. First, the teacher always assigns $R^*(1)$ above $c$ and never assigns $R^*(1) = c_-$. Hence, a \textit{bunch} (shaded area) in the observed density $f_R(r)$ appears. Second, the teacher always selects students to manipulate only from the failed students $R^*(0) < c$ and never manipulates passing students $R^*(0) \geq c$. Hence, a \textit{notch} (dotted area) appears at the cutoff point. Consequently, the realized score density, $f_R(r)$, jumps by $f_R(c_+) - f_R(c_-)$, indicating the presence of manipulation.

\begin{figure}
    \centering
    \ifnum\submissionver=1%
    \includegraphics[width=\columnwidth]{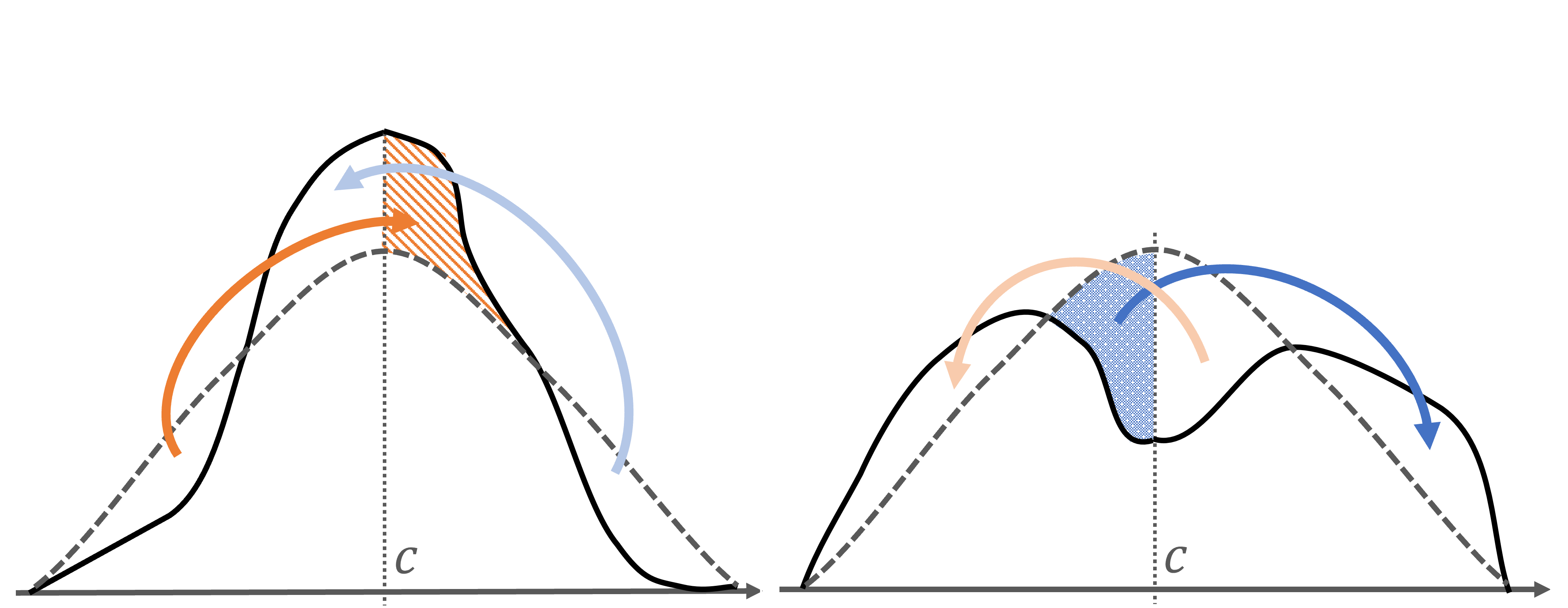}%
    \fi
    \caption{Examples of detection failures.}
    \label{fig:detectfail}
    \begin{figurenotes}
    The dashed line represents the density of $R^*(0)$; the solid line represents the density of $R$, $f_{R}(\cdot)$. Detection fails by bunching (left) at and notching (right) from both sides.
    \end{figurenotes}
\end{figure}

However, detection fails if bunches or notches are present on the other side, as illustrated in Figure \ref{fig:detectfail}. If a teacher assigns $R^*(1)$ for the control $R^*(1) < c$ to intentionally fail someone (Figure \ref{fig:detectfail} left), then the bunches may cancel out. Similarly, if another teacher selects $R^*(0)$ from the treated $R^*(0) \geq c$ to fail them (Figure \ref{fig:detectfail} right), then the notches may cancel out. In other words, the detection strategy can fail with \textit{two-sided} incentives for manipulation; some favor treatment, while others are against it. An auxiliary assumption is necessary to prevent such coincidental failure.

Now we introduce the auxiliary assumption formally. We first assume that $R^*(0)$ is ex-ante randomized locally. This assumption is virtually a definition for $R^*(0)$ and $M$: $M$ indicates manipulation that violates either \eqref{eq.control} or \eqref{eq.selection}, and $R^*(0)$ is the running variable after incurring all other innocuous manipulations that satisfy both \eqref{eq.control} and \eqref{eq.selection}.
\begin{assumption} \label{ass.exanteFreeNoG}
\begin{eqnarray*}
    & &  f_{R^*(0)}(c_+) = f_{R^*(0)}(c_-) \\
    & & \hspace{0.3in} \mbox{ and } E[Y(d)|R^*(0)=c_-] = E[Y(d)|R^*(0)=c_+], \ d \in \{0,1\}.
\end{eqnarray*}
\end{assumption}

The auxiliary assumption imposes two restrictions: Manipulation $M$ always assigns treatment whenever \eqref{eq.control} fails and never selects units from the treated whenever \eqref{eq.selection} fails. These are \textit{one-sided} manipulations because they always favor the treatment and never the control side.
\begin{assumption}[One-sided manipulation] \label{ass.one-sided} 
Either of the following holds:
\begin{eqnarray*}
    & & \mbox{(i) } P(R^*(1) \geq c| M = 1) = 1, \mbox{ and } P(M = 0|R^*(0) \geq c) = 1,\\
    & & \mbox{(ii) } P(R^*(1) \geq c| M = 1) = 1, \mbox{ and \eqref{eq.selection} holds, or }\\
    & & \mbox{(iii) } \mbox{\eqref{eq.control} holds, and } P(M = 0|R^*(0) \geq c) = 1.
\end{eqnarray*}
\end{assumption}

Under the auxiliary assumption, continuous density implies identification. Hence, passing the density test confirms the identification. \footnote{In general, any mixture of three types in Assumption \ref{ass.one-sided} is allowed with complex notations. See Appendix for the general cases.}
\begin{theorem} \label{thm.main}
    Under Assumption \ref{ass.exanteFreeNoG} and \ref{ass.one-sided}, if $f_{R}(c_-) = f_{R}(c_+)$, \eqref{eq.HTV} holds.
\end{theorem}

\begin{proof}
 The proof is in the Appendix.
\end{proof}

Theorem \ref{thm.main} also justifies the balance or placebo test; namely, the conditional means of covariates are continuous at the cutoff under the null hypothesis of the density test:
\begin{corollary} \label{cor.balance}
Under Assumption \ref{ass.exanteFreeNoG} and \ref{ass.one-sided}, replacing $Y(d)$ with a pre-determined covariate $W$, $f_R(c_-) = f_R(c_+)$ implies $E[W|R=c_+] = E[W|R=c_-]$.
\end{corollary}

\begin{remark} \label{rem.grr}
    Assumption \ref{ass.one-sided} shares the same name as a restriction in \cite{gerardBoundsTreatmentEffects2020} for partial identification. They consider an indicator $M^{GRR} \in \{0,1\}$ for a subset of units that are \textit{always-assigned}, whereas our $M$ indicates the \textit{action} of manipulation. They impose high-level conditions such that \eqref{eq.localrandomized} holds, conditional on $M^{GRR} = 0$ and $P(R \geq c| M^{GRR}=1) = 1$, namely, local randomization holds when we remove the always-assigned $\{M^{GRR} = 1\}$ that are present only in the right of the cutoff. With a particular manipulation in mind, one needs to construct $M^{GRR}$ such that $\{M^{GRR} = 0\}$ satisfies local randomization. It can be complicated to construct $M^{GRR}$ from a particular action of manipulation, for example, as in Case (iii) in Assumption \ref{ass.one-sided}. Appendix reveals that our low-level conditions can generate $M^{GRR}$, which satisfies their restriction, from $M$ under our restrictions.
\end{remark}

\section{Discussions for Empirical Practices} \label{sec.discussion}

We have provided two main results clarifying new identification conditions (Proposition \ref{prop.identify}) and conditions for the density and balance tests being valid (Theorem \ref{thm.main} and Corollary \ref{cor.balance}). These conditions are missing or implicit in existing empirical studies. We revisit a few empirical studies to discuss their implications for practices.

\subsection{Identifying arguments in empirical studies}

From Proposition \ref{prop.identify}, identification holds if manipulation randomly assigns $R^*(1)$ around the cutoff and randomly selects units whose $R^*(0)$ are around the cutoff. The former can be justified by the famous \textit{no precise control} or \textit{imprecise control} strategy that ``individuals do not precisely manipulate $X$ around the threshold has the prediction that treatment is locally randomized.'' \citep[page. 295, $X$ is the running variable]{leeRegressionDiscontinuityDesigns2010}. A similar claim is seen in the latest textbook: ``if units lack the ability to precisely manipulate the score value they receive, there should be no systematic differences between units with similar values of the score.'' \citep[page. 79]{Cattaneo_Idrobo_Titiunik_2024}. 

This \textit{no precise control} strategy is widely used for identification in empirical studies. For example, \cite{Dahl_Loken_Mogstad_2014} study the impact of eligibility for paternal parental leave on his peer fathers in the workplace. They state that ``The key identifying assumption of our fuzzy RD design is that individuals are unable to precisely control the assignment variable, date of birth, near the cutoff date c, in which case the variation in treatment near c is random.'' \cite{Pinotti_2017} studies the impact of a work permit on crime rates by exploiting the permit being offered when the application is submitted before an unknown threshold time. He claims that ``These complexities provide a compelling argument for the fundamental identification assumption that applicants within an arbitrarily narrow bandwidth of the cutoff were unable to precisely determine their assignment to either side of it.''. Similar arguments appear in more recent articles such as \citet{dechezlepretreTaxIncentivesIncrease2023} and \cite{huangPoliticalInfluenceBank2024}. However, they are insufficient because no precise control implies the inability to control the manipulated $R^*(1)|M=1$ but remains silent about the non-manipulated $R^*(0)|M=0$. For the continuity condition \citep{hahnIdentificationEstimationTreatment2001}, we need to regulate who would be manipulated in terms of their \textit{non-manipulated} values, $R^*(0)$. Hence, their identification discussions may be incomplete.

Failure in the latter condition on $R^*(0)|M=0$ is addressed in some empirical studies, although its mechanism is under-explored. \cite{jepsenLaborMarketReturns2016} document a discontinuous density function for a design with the test score as the running variable, blaming a selection due to examination retakes. In Example \ref{ex.almostnec2}, we detailed two mechanisms in their context. An obvious mechanism is the imbalance in mean outcomes by self-selection due to manipulation. The other is non-trivial: imbalanced densities can fail identification without self-selection in their outcomes.

Our framework is simple and applicable to most RD designs. For example, in \cite{Dahl_Loken_Mogstad_2014}, if a father quits a job because he is ineligible for the program, he would have no workplace and associated dependent variable. Such an attrition of the dependent variable is a threat for the local randomization of $R^*(0)|M=0$ because ineligible fathers are more likely to have $R^*(0)$ missing than others. A similar consideration justifies the identification in \cite{Pinotti_2017}. If \cite{Pinotti_2017} had the universe of the work permit application timestamps and no systematic selection is possible in terms of $R^*(0)|M=0$, the identification should hold. Thus, our procedure offers a sophisticated check for those identification concerns in any design. 

\subsection{Illustrations of the one-sided restriction failures} \label{sec.when_shall}

When manipulation can violate either \eqref{eq.control} or \eqref{eq.selection}, diagnostic tests should be employed to detect such manipulation, for which the condition for Theorem \ref{thm.main} must hold. 
From Theorem \ref{thm.main}, we must verify that any manipulation of $R^*(1)|M=1$ must favor $R^*(1) \geq c$ whenever it fails \eqref{eq.control} and any selection of $R^*(0)$ should be from $R^*(0) < c$ whenever it fails \eqref{eq.selection}. The former is straightforward: if someone is willing and able to assign their $R^*(1)$ just above the cutoff, no one must be willing or able to assign their $R^*(1)$ just below it.

The latter can be critical for some designs. \cite{bradley_unions_2017} study the impact of labor union (National Labor Relations Board) formation in U.S. firms on patent-related variables. Many studies use union voting as a device for an RD design. \cite{dinardoEconomicImpactsNew2004} describe a typical process for a U.S. union formation. Among their described procedures, the followign \textit{Step 5.} suggests that a re-election can be proposed via objection to the initial election.
\begin{quoting}
     5. Within seven days after the final tally of the ballots, parties can file objections to how the election was conducted. With sufficient evidence that the election was not carried out properly, the NLRB can rule to invalidate the outcome of an election, and conduct another one thereafter. Specific ballots cannot be challenged after the voting is completed.\citep[page. 1389]{dinardoEconomicImpactsNew2004}
\end{quoting}

The re-election is manipulation, although it may sound as innocuous because no one can ensure the final election results via re-election. However, re-elections violate the condition for $R^*(0)|M=0$ \eqref{eq.selection} because objections would be reasonably made only from the failing side that can be for or against the union. 

Many designs should be able to detect manipulation. For example, \citet{Angrist_Lavy_Leder-Luis_Shany_2019} is a follow-up study of \cite{Angrist_Lavy_1999} that exploits the \textit{Maimonides' rule} in the Israeli school system: a school with $41$ students must have two classes of $20$ and $21$. The treatment is the assignment to a smaller class, but schools may have manipulated the enrollment as ``schools are warned not to move students between grades or to enroll those overseas so as to produce an additional class'' \citep{Angrist_Lavy_Leder-Luis_Shany_2019} by the Israeli Ministry of Education (MOE). Such manipulation is costly and should occur only to exceed the cutoff because ``School leaders might care to do this because educators and parents prefer smaller classes. MOE rules that set school budgets as an increasing function of the number of classes also reward manipulation'' (\citealp{Angrist_Lavy_Leder-Luis_Shany_2019}). Hence, the density test is valid for their design. \cite{otsuEstimationInferenceDiscontinuity2013} and \cite{Angrist_Lavy_Leder-Luis_Shany_2019} report the discontinuity of the density in \cite{Angrist_Lavy_1999} data. While continuous density is sufficient, it is not necessary for identification. \cite{Angrist_Lavy_Leder-Luis_Shany_2019} verify that the index of socioeconomic status is unrelated to Maimonides's rule conditional on a few covariates, and such logic may justify identification. \footnote{As in \eqref{eq.decomp} and example \ref{ex.almostnec2}, identification failure is driven by a difference in manipulators and non-manipulators. If $E[Y(d)|R^*(m)=c_-,M=m] = E[Y(d)|R^*(m)=c_+,M=m], m \in \{0,1\}$ and these means are the same across $m \in \{0,1\}$, then identification holds with discontinuous densities.} It may also be consistent with \cite{Arai_Hsu_Kitagawa_Mourifie_Wan_2021} who report the passage of their fuzzy RD test, which does not involve any restrictions on the density but says nothing about its reduced-form sharp design. Continuous density confirms identification, but designs with discontinuous density may also be salvaged, despite being challenging.

\section{Conclusion}\label{sec.conc}

RD identification is valid for an ideal design that assigns treatment as if it is randomized locally at the cutoff. However, individuals may manipulate the running variable. Because manipulation has never been defined in existing models, existing identification is based on high-level conditions that are silent about \textit{when a design is ideal under manipulation}.

We provide the low-level condition for identification by introducing the potential outcome framework for manipulation. In Section \ref{sec.identification}, we derive simple low-level conditions for identification as restrictions on manipulation via the framework. Low-level conditions require that manipulation to be randomization, which can be achieved by prohibiting two \textit{precise} manipulations: the manipulated running variable must not be precisely assigned on a particular side of the cutoff and units for manipulation must not be precisely selected from a particular side of the cutoff. These restrictions arise from a decomposition of the continuity condition \citep{hahnIdentificationEstimationTreatment2001} for identification; the continuity condition follows from the balanced mean potential outcomes weighted by the balanced densities. This decomposition highlights the critical role of continuous density in our framework.

Furthermore, in Section \ref{sec.mrrd}, we established the low-level auxiliary assumption that guarantees that diagnostic tests can detect manipulations. Under the proposed auxiliary assumption, we reveal that the null hypothesis of the popular density test \citep{mccraryManipulationRunningVariable2008} implies identification. The auxiliary assumption eliminates manipulation with two-sided incentives. Our restriction is explicit in manipulation as an action to influence the running variable. Hence, the auxiliary assumption can be justified from particular stories about manipulation in consideration.

In Section \ref{sec.discussion}, we discuss the consequences of our framework in published empirical studies, documenting that identification claims in previous studies may be incomplete and providing remedies against it. Furthermore, we highlight a study that may not satisfy the auxiliary assumption and hence may be incapable of detecting manipulation. The possible detection failure is suggested by a two-sided incentive for initiating manipulation that leads to precise selection from either side of the cutoff. We emphasize that the manipulation, which sounds innocuous in existing models, can be a threat to its detection and identification.

The study has some limitations. First, the distribution of $R^*(0)$ or its proxy variable may be available; however, this additional information may improve testing and identification. Second, the predetermined covariates may have alternative uses based on our analysis. Many studies propose estimating with covariates: \cite{Frolich_Huber_2019} propose a method with a multi-dimensional non-parametric estimation; \citet{Calonico_Cattaneo_Farrell_Titiunik_2019} develop an easy-to-implement augmentation; \citet{Noack_Olma_Rothe_2021} consider flexible and efficient estimation including machine-learning devices; \cite{Kreiss_Rothe_2022} and \citet{Arai_Otsu_Seo_2021} explore augmentation with high-dimensional covariates. Nevertheless, covariates are rarely used to adjust for a possible identification failure and our results may suggest an alternative use of covariates.
Finally, our analysis may not be trivial with a multi-dimensional running variable $R$. It would also be promising to extend our analysis to RD designs with multiple cutoff values for which \citet{Cattaneo_Keele_Titiunik_Vazquez-Bare_2016} propose a pooling parameter and its implementation and \citet{Cattaneo_Keele_Titiunik_Vazquez-Bare_2021} consider an extrapolation method. Developing conceptual and practical recommendations for these designs is a future issue to be explored. 

\newpage

\appendix

\section{Appendix. Regularity conditions and proofs} 
\label{apx.A}

\begin{assumption}[Regularity conditions]\label{ass.regularnoG} 
(i) For $m \in \{0,1\}$, the conditional density function $f_{R^{\ast}(m)|M=m}(r)$ exists and has left and right limits, that is 
\[
f_{R^{\ast}(m)|M=m}(c_{-}) \equiv \lim_{r \uparrow c} f_{R^{\ast}(m)|M=m}(r) \mbox{ and } f_{R^{\ast}(m)|M=m}(c_{+}) \equiv \lim_{r \downarrow c} f_{R^{\ast}(m)|M=m}(r).
\] (ii) For $m \in \{0,1\}$, the conditional probability $P(M=m|R^{\ast}(m)=r)$ exists and has left and right limits, that is 
\begin{align*}
&P(M=m|R^{\ast}(m)=c_{-}) \equiv \lim_{r \uparrow c} P(M=m|R^{\ast}(m)=r), \mbox{ and }\\
&P(M=m|R^{\ast}(m)=c_{+}) \equiv \lim_{r \downarrow c} P(M=m|R^{\ast}(m)=r).    
\end{align*}
(iii) For $d \in \{0,1\}$ and $m \in \{0,1\}$, the conditional expectation $E[Y(d)|R^{\ast}(m)=r, M=m]$ exists and has left and right limits, that is 
\begin{align*}
    &E[Y(d)|R^{\ast}(m)=c_{-}, M=m] \equiv \lim_{r \uparrow c} E[Y(d)|R^{\ast}(m)=r, M=m] \mbox{ and }\\
    &E[Y(d)|R^{\ast}(m)=c_{+}, M=m] \equiv \lim_{r \downarrow c} E[Y(d)|R^{\ast}(m)=r, M=m].
\end{align*}
\end{assumption}
\begin{proof}[Proof of Theorem \ref{thm.main}]
We observe that
\begin{eqnarray*}
f_{R}(r) &=& f_{R^{\ast}(1)|M=1}(r) P(M=1) + f_{R^{\ast}(0)|M=0}(r) P(M=0) \\
&=& f_{R^{\ast}(1)|M=1}(r) P(M=1) + P(M=0|R^{\ast}(0) = r) f_{R^{\ast}(0)}(r).
\end{eqnarray*}
If Assumption \ref{ass.one-sided} (i) holds, we obtain
\begin{eqnarray}
f_{R}(c_{+}) - f_{R}(c-) &=& f_{R^{\ast}(1)|M=1}(c_{+}) P(M=1) \nonumber \\
& & \hspace{0.3in} + \left\{ 1 - P(M=0|R^{\ast}(0) = c_{-})\right\} f_{R^{\ast}(0)}(c), \nonumber
\end{eqnarray}
which implies $f_{R}(c_{-}) \leq f_{R}(c_{+})$. Hence, $f_{R^{\ast}(1)|M=1}(c_{+})P(M=1) = 0$ and $P(M=0|R^{\ast}(0) = c_{-}) = 1$ hold if $f_{R}(c_{-}) = f_{R}(c_{+})$. Because
\[
f_{R^{\ast}(1)|M=1}(c_{-})P(M=1) = f_{R^{\ast}(1)|M=1}(c_{+})P(M=1) = 0,
\]
we have
\begin{eqnarray*}
& & E[Y(d)|R^{\ast}(1)=c_{-},M=1] f_{R^{\ast}(1)|M=1}(c_{-}) P(M=1) \\
& & \hspace{0.5in} = E[Y(d)|R^{\ast}(1)=c_{+},M=1] f_{R^{\ast}(1)|M=1}(c_{+})P(M=1).
\end{eqnarray*}
Additionally, because $P(M=0|R^{\ast}(0) = c_{-}) = P(M=0|R^{\ast}(0) = c_{+}) = 1$, for $r = c_{-}, c_{+}$, we have
\begin{eqnarray*}
f_{R^{\ast}(0)| M=0}(r) P(M=0) &=& P(M=0|R^{\ast}(0)=r) f_{R^{\ast}(0)}(c) \ = \ f_{R^{\ast}(0)}(c), \\
E[Y(d)|R^{\ast}(0)=r, M=0] &=& E[Y(d)|R^{\ast}(0)=c].
\end{eqnarray*}
Hence, from (\ref{eq.decomp}) it follows that $E[Y(d)|R=r]$ is continuous at $r=c$.

Next, suppose that Assumption \ref{ass.one-sided} (ii) holds. Then, 
\[
f_{R}(c_{+}) - f_{R}(c-) \ = \ f_{R^{\ast}(1)|M=1}(c_{+}) P(M=1).
\]
This implies that $f_{R^{\ast}(1)|M=1}(c_{+}) P(M=1) = 0$ holds if $f_{R}(c_{-}) = f_{R}(c_{+})$. Following the same argument as above, we can demonstrate the continuity of $E[Y(d)|R=r]$ at $r=c$.

Finally, suppose that Assumption \ref{ass.one-sided} (iii) holds. Then 
\[
f_{R}(c_{+}) - f_{R}(c-) \ = \ \left\{ 1 - P(M=0|R^{\ast}(0) = c_{-})\right\} f_{R^{\ast}(0)}(c).
\]
This implies that $P(M=0|R^{\ast}(0) = c_{-})=1$ holds if $f_{R}(c_{-}) = f_{R}(c_{+})$. Following the same argument as above, we can demonstrate the continuity of $E[Y(d)|R=r]$ at $r=c$.
\end{proof}

\section{Appendix. General case}  \label{apx.general}

\begin{assumption}[Regularity conditions]\label{ass.general} 
Let $G \in \{0,1,2,3\}$ be an unobserved type indicator that satisfies the following conditions. (i) For $g \in \{0,1,2,3\}$, Assumptions \ref{ass.regularnoG} and \ref{ass.exanteFreeNoG} hold conditional on $G=g$. (ii) Conditional on $G=0$, (\ref{eq.control}) and (\ref{eq.selection}) hold. (iii) Conditional on $G=1$, Assumption \ref{ass.one-sided} (i) holds. (iv) Conditional on $G=2$, Assumption \ref{ass.one-sided} (ii) holds. (iii) Conditional on $G=3$, Assumption \ref{ass.one-sided} (iii) holds.
\end{assumption}

\begin{theorem} \label{thm.general}
Under Assumption \ref{ass.general}, if $f_{R}(c_-) = f_{R}(c_+)$ then \eqref{eq.HTV} holds; hence, the ATE is identified.
\end{theorem}

\begin{proof}
By proofs of Proposition \ref{prop.identify} and Theorem \ref{thm.main}, we obtain $f_{R|G=g}(c_{-}) \leq f_{R|G=g}(c_{+})$ and $f_{R|G=g}(c_{-}) = f_{R|G=g}(c_{+})$ implies $E[Y(d)|R=c_{-},G=g] = E[Y(d)|R=c_{+},G=g]$. Because we have $f_{R|G=g}(c_{-}) \leq f_{R|G=g}(c_{+})$ and
\[
f_{R}(r) = \sum_{g=0}^3 f_{R|G=g}(r),
\]
$f_{R}(c_{-}) = f_{R}(c_{+})$ implies that $f_{R|G=g}(c_{-}) = f_{R|G=g}(c_{+})$ holds for all $g$. Hence, if $f_{R}(c_{-}) = f_{R}(c_{+})$, we obtain $f_{R|G=g}(c_{-}) = f_{R|G=g}(c_{+})$ and $E[Y(d)|R=c_{-},G=g] = E[Y(d)|R=c_{+},G=g]$ for all $g$. Additionally, we observe that
\begin{eqnarray*}
E[Y(d)|R=r] &=& \sum_{g=0}^3 E[Y(d)|R=r,G=g] P(G=g|R=r) \\
&=& \sum_{g=0}^3 E[Y(d)|R=r,G=g] \cdot \frac{f_{R|G=g}(r) P(G=g)}{f_{R}(r)}.
\end{eqnarray*}
As a result, if $f_{R}(c_{-}) = f_{R}(c_{+})$, $E[Y(d)|R=r]$ is continuous at $r=c$.
\end{proof}

\section{Appendix. Comparison with Gerard et al. (2020)}  \label{apx.GRR}

In this section, we demonstrate that the assumptions of \cite{gerardBoundsTreatmentEffects2020} are satisfied under Assumptions \ref{ass.exanteFreeNoG}--\ref{ass.one-sided} if our assumptions regarding the conditional expectations of $Y(d)$ are replaced by those regarding conditional distributions of $Y(d)$. \cite{gerardBoundsTreatmentEffects2020} assume that there exists an indicator $M^{GRR} \in \{0,1\}$ such that
\begin{eqnarray}
& & F_{Y(d)|R, M^{GRR}=0}(y|c_{-}) \ = \ F_{Y(d)|R, M^{GRR}=0}(y|c_{+}), \label{GRR_1} \\
& & P(R \geq c | M^{GRR}=1) \ = \ 1, \label{GRR_2}
\end{eqnarray} 
where $F_{U|V}$ denotes the conditional distribution function of $U$ conditional on $V$ for any random variables $U$ and $V$. We reveal that we can construct $M^{GRR}$ satisfying (\ref{GRR_1}) and (\ref{GRR_2}) under Assumptions \ref{ass.exanteFreeNoG}--\ref{ass.one-sided}. In the following, we replace our assumptions regarding the conditional expectations with those regarding conditional distributions. For example, the continuity of $E[Y(d)|R^{\ast}(0) = r]$ in Assumption \ref{ass.exanteFreeNoG} is replaced by
\[
F_{Y(d)|R^{\ast}(0)}(y|c_{-}) \ = \ F_{Y(d)|R^{\ast}(0)}(y|c_{+}) \ \text{for any $y$.}
\]

First, we demonstrate that $M^{GRR} \equiv M$ satisfies (\ref{GRR_1}) and (\ref{GRR_2}) under Assumption \ref{ass.one-sided} (ii). By Assumption \ref{ass.one-sided} (ii), we obtain
\begin{eqnarray*}
F_{Y(d)|R, M^{GRR}=0}(y|c_{-}) &=& F_{Y(d)|R^{\ast}(0), M=0}(y|c_{-}) \\
&=& F_{Y(d)|R^{\ast}(0), M=0}(y|c_{+}) \ = \ F_{Y(d)|R, M^{GRR}=0}(y|c_{+}), \\
P(R \geq c | M^{GRR}=1) &=& P(R^{\ast}(1) \geq c | M=1) \ = \ 1.
\end{eqnarray*}
Hence, under Assumption \ref{ass.one-sided} (ii), $M$ plays the same role of $M^{GRR}$ in \cite{gerardBoundsTreatmentEffects2020}.

Second, we consider Assumption \ref{ass.one-sided} (i). Because $f_{R^{\ast}(0)}(r)$ and $F_{Y(d)|R^{\ast}(0)}(y|r)$ are continuous at $r=c$, we can construct a group indicator $M^{\ast} \in \{0,1\}$\footnote{For simplicity, we assume that conditional density functions exist $f_{Y(d)|R^{\ast}(0)}(y|c_{-})$, $f_{Y(d)|R^{\ast}(0),M=0}(y|c_{-})$, and $f_{Y(d)|R^{\ast}(0),M=1}(y|c_{-})$. Letting
\begin{eqnarray*}
& & g(y) \equiv f_{Y(d)|R^{\ast}(0)}(y|c_{-}), \ \ g_0(y) \equiv f_{Y(d)|R^{\ast}(0),M=0}(y|c_{-}), \\
& & g_1(y) \equiv f_{Y(d)|R^{\ast}(0),M=1}(y|c_{-}), \ \ \text{and} \ \ p_M \equiv P(M=1|R^{\ast}(0)=c_{-}),
\end{eqnarray*}
then $g(y)$ can be written as $g(y) = p_M g_1(y) + (1-p_M) g_0(y)$. Let $U$ be a uniformly distributed random variable independent of other variables. Define $M^{\ast} \equiv 1 \left\{ U \leq \frac{p_Mg_1(Y(d))}{g(Y(d))} \ \text{and} \ R^{\ast}(0) \geq c \right\}$. Then, $P(R^{\ast}(0) \geq c | M^{\ast} =1)=1$ holds by the definition of $M^{\ast}$. Because we have $f_{Y(d)|R^{\ast}(0)}(y|c_{+}) = f_{Y(d)|R^{\ast}(0)}(y|c_{-}) = g(y)$, we obtain
\begin{eqnarray*}
P(M^{\ast}=1|R^{\ast}(0)=c_{+}) &=& E\left[ \frac{p_Mg_1(Y(d))}{g(Y(d))} \Big| R^{\ast}(0) = c_{+}\right] \\
&=& \int \left( \frac{p_Mg_1(y)}{g(y)} \right) g(y) dy \ = \ P(M=1|R^{\ast}(0)=c_{-}).
\end{eqnarray*}
Furthermore, using Bayes's theorem, we obtain
\begin{eqnarray*}
f_{Y(d)|R^{\ast}(0),M^{\ast}=1}(y|c_{+}) &=& \frac{P(M^{\ast}=1|Y(d)=y, R^{\ast}(0) = c_{+}) f_{Y(d)|R^{\ast}(0)}(y|c_{+})}{P(M^{\ast}=1|R^{\ast}(0)=c_{+})} \\
&=&\frac{\{ p_M g_1(y) / g(y) \} \cdot g(y)}{p_M} \ = \ f_{Y(d)|R^{\ast}(0),M=1}(y|c_{-}),
\end{eqnarray*}
which implies $F_{Y(d)|R^{\ast}(0),M^{\ast}=1}(y|c_{+}) = F_{Y(d)|R^{\ast}(0),M=1}(y|c_{-})$.} such that
\begin{eqnarray}
F_{Y(d)|R^{\ast}(0),M^{\ast}=1}(y|c_{+}) &=& F_{Y(d)|R^{\ast}(0),M=1}(y|c_{-}), \label{M_star_1} \\
P(M^{\ast}=1|R^{\ast}(0)=c_{+}) &=& P(M=1|R^{\ast}(0)=c_{-}), \label{M_star_2} \\
P(R^{\ast}(0) \geq c | M^{\ast}=1 ) &=& 1. \label{M_ster_3}
\end{eqnarray}
Then, we can demonstrate that $M^{GRR} \equiv 1\{M=1 \ \text{or} \ M^{\ast}=1\}$ satisfies (\ref{GRR_1}) and (\ref{GRR_2}). Because $P(R^{\ast}(1) \geq c | M=1)=1$ and $P(R^{\ast}(0) \geq c | M^{\ast}=1)=1$, we have
\begin{eqnarray*}
& & P(R \geq c | M^{GRR} = 1) \\
&=& P(R \geq c | M=1 \ \text{or} \ M^{\ast}=1) \\
&=& \frac{P(R^{\ast}(1) \geq c, M=1) + P(R^{\ast}(0) \geq c, M=0, M^{\ast}=1)}{P(M=1 \ \text{or} \ M^{\ast}=1)} \\
&=& \frac{P(M=1) + P(M=0, M^{\ast}=1)}{P(M=1 \ \text{or} \ M^{\ast}=1)} \ = \ 1.
\end{eqnarray*}
Because $\{M^{GRR} =0\} = \{M=0 \ \text{and} \ M^{\ast}=0\}$ and $P(M=0|R^{\ast}(0)=c_{+})=1$, we have
\begin{eqnarray*}
F_{Y(d)|R, M^{GRR}=0}(y|c_{+}) &=& F_{Y(d)|R^{\ast}(0), M^{\ast}=0}(y|c_{+}), \\
P(M^{GRR}=0|R^{\ast}(0)=c_{+}) &=& P(M^{\ast}=0|R^{\ast}(0)=c_{+}).
\end{eqnarray*}
Furthermore, it follows from $P(R^{\ast}(0) < c, M^{\ast}=1)=0$ that we have
\begin{eqnarray*}
P(M^{\ast}=0|R^{\ast}(0) < c) &=& \frac{P(R^{\ast}(0) < c, M^{\ast}=0)}{P(R^{\ast}(0) < c)} \\
&=& \frac{P(R^{\ast}(0) < c)-P(R^{\ast}(0) < c, M^{\ast}=1)}{P(R^{\ast}(0) < c)} \ = \ 1.
\end{eqnarray*}
Hence, we obtain
\begin{eqnarray*}
F_{Y(d)|R, M^{GRR}=0}(y|c_{+}) &=& \left( \frac{1}{P(M^{\ast}=0|R^{\ast}(0)=c_{+})} \right) \Big\{ F_{Y(d)|R^{\ast}(0)}(y|c_{+}) \\
& & \hspace{0.1in} - P(M^{\ast}=1|R^{\ast}(0)=c_{+}) F_{Y(d)|R^{\ast}(0), M^{\ast}=1}(y|c_{+}) \Big\} \\
&=& \left( \frac{1}{P(M=0|R^{\ast}(0)=c_{+})} \right) \Big\{ F_{Y(d)|R^{\ast}(0)}(y|c_{-}) \\
& & \hspace{0.1in} - P(M=1|R^{\ast}(0)=c_{-}) F_{Y(d)|R^{\ast}(0), M=1}(y|c_{-}) \Big\} \\
&=& F_{Y(d)|R^{\ast}(0), M=0}(y|c_{-}) \ = \ F_{Y(d)|R, M^{GRR}=0}(y|c_{-}),
\end{eqnarray*}
where the last equality follows because $P(M^{\ast}=0|R^{\ast}(0) < c) = 1$. Therefore, $M^{GRR}$ satisfies (\ref{GRR_1}) and (\ref{GRR_2}).

Finally, we consider Assumption \ref{ass.one-sided} (iii). Similar to the above case, we construct $M^{\ast} \in \{0,1\}$ satisfying (\ref{M_star_1}), (\ref{M_star_2}), and (\ref{M_ster_3}). Then, we demonstrate that $M^{GRR} \equiv 1 \{ M=0 \ \text{and} \ M^{\ast}=1 \}$ satisfies (\ref{GRR_1}) and (\ref{GRR_2}). Because $P(R^{\ast}(0) \geq c | M^{\ast}=1 )=1$ and $P(M=0|R^{\ast}(0) \geq c)=1$, we have $P(M=0|M^{\ast} = 1) = 1$. Hence, we obtain
\begin{eqnarray*}
P(R \geq c | M^{GRR}=1) &=& P(R^{\ast}(0) \geq c | M=0, M^{\ast}=1) \\
&=& P(R^{\ast}(0) \geq c | M^{\ast}=1) \ = \ 1,
\end{eqnarray*}
where the last equality follows from $P(R^{\ast}(0) \geq c | M^{\ast}=1 )=1$. Therefore, (\ref{GRR_2}) is satisfied.

Next, we demonstrate that $M^{GRR}$ satisfies (\ref{GRR_1}). Because $\{M^{GRR} = 0\} = \{M=1 \ \text{or} \ M^{\ast} = 0\}$, we observe that
\begin{eqnarray}
F_{Y(d)|R, M^{GRR}=0}(y|c_{+}) &=& P(M=0|R=c_{+}, M^{GRR}=0) \nonumber \\
& & \hspace{0.5in} \times F_{Y(d)|R^{\ast}(0), M = 0, M^{\ast}=0}(y|c_{+})\nonumber \\
& &  + P(M=1|R=c_{+}, M^{GRR}=0) \nonumber \\
& & \hspace{0.5in} \times F_{Y(d)|R^{\ast}(1), M = 1}(y|c_{+}), \label{GRR_case3_1} 
\end{eqnarray}
where $F_{Y(d)|R^{\ast}(1), M = 1}(y|c_{+}) = F_{Y(d)|R^{\ast}(1), M = 1}(y|c_{-})$ holds by Assumption \ref{ass.one-sided} (iii). Because $M^{\ast}$ satisfies (\ref{M_star_1}) and (\ref{M_star_2}), we observe that
\begin{eqnarray}
& & F_{Y(d)|R^{\ast}(0), M = 0, M^{\ast}=0}(y|c_{+}) \nonumber \\
&=& F_{Y(d)|R^{\ast}(0), M^{\ast}=0}(y|c_{+}) \nonumber \\
&=& \left( \frac{1}{P(M^{\ast}=0|R^{\ast}(0)=c_{+})} \right) \Big\{ F_{Y(d)|R^{\ast}(0)}(y|c_{+}) \nonumber \\
& & \hspace{0.8in} - P(M^{\ast}=1|R^{\ast}(0)=c_{+})  F_{Y(d)|R^{\ast}(0),M^{\ast}=1}(c_{+})  \Big\} \nonumber \\
&=& \left( \frac{1}{1-P(M=1|R^{\ast}(0)=c_{-})} \right) \Big\{ F_{Y(d)|R^{\ast}(0)}(y|c_{-}) \nonumber \\
& & \hspace{0.8in} - P(M=1|R^{\ast}(0)=c_{-}) F_{Y(d)|R^{\ast}(0),M=1}(y|c_{-}) \Big\} \nonumber \\
&=& F_{Y(d)|R^{\ast}(0), M=0}(y|c_{-}) \ = \ F_{Y(d)|R^{\ast}(0), M=0, M^{\ast}=0}(y|c_{-}), \label{GRR_case3_2}
\end{eqnarray}
where the first equality follows from $P(M=0|R^{\ast}(0) \geq c) = 1$ and the last equality follows because $P(R^{\ast}(0) \geq c | M^{\ast}=1 ) = 1$ implies $P(M^{\ast}=0 | R^{\ast}(0) < c) = 1$. We observe that
\begin{eqnarray}
& & P(M=1|R=c_{+}, M^{GRR}=0) \nonumber \\
&=& \frac{P(M=1|R=c_{+})}{P(M=1|R=c_{+}) + P(M=0, M^{\ast} = 0|R=c_{+})} \nonumber \\
&=& \frac{f_{R^{\ast}(1)|M=1}(c_{+}) P(M=1)}{f_{R^{\ast}(1)|M=1}(c_{+}) P(M=1) + f_{R^{\ast}(0)|M=0,M^{\ast}=0}(c_{+}) P(M=0,M^{\ast}=0)} \nonumber \\
&=& \frac{f_{R^{\ast}(1)|M=1}(c_{+}) P(M=1)}{f_{R^{\ast}(1)|M=1}(c_{+}) P(M=1) + P(M=0, M^{\ast} = 0|R^{\ast}(0)=c_{+})f_{R^{\ast}(0)}(c_{+})}, \nonumber 
\end{eqnarray}
where $f_{R^{\ast}(1)|M=1}(c_{+}) = f_{R^{\ast}(1)|M=1}(c_{-})$ and $f_{R^{\ast}(0)}(c_{+}) = f_{R^{\ast}(0)}(c_{-})$ hold. Additionally, because $P(M=0|R^{\ast}(0)=c_{+})=1$, $P(M^{\ast}=0|R^{\ast}(0)=c_{+})=P(M=0|R^{\ast}(0)=c_{-})$, and $P(M^{\ast}=0 | R^{\ast}(0)=c_{-}) = 1$, we obtain
\begin{eqnarray*}
P(M=0, M^{\ast} = 0|R^{\ast}(0)=c_{+}) &=& P(M^{\ast} = 0|R^{\ast}(0)=c_{+}) \\
&=& P(M = 0|R^{\ast}(0)=c_{-}) \\
&=& P(M=0, M^{\ast} = 0|R^{\ast}(0)=c_{-}).
\end{eqnarray*}
Therefore, we obtain
\begin{equation}
P(M=1|R=c_{+}, M^{GRR}=0) \ = \ P(M=1|R=c_{-}, M^{GRR}=0). \label{GRR_case3_3}
\end{equation}
Thus, by substituting (\ref{GRR_case3_2}) and (\ref{GRR_case3_3}) for (\ref{GRR_case3_1}), we obtain (\ref{GRR_1}).

\newpage

\bibliographystyle{ecta}
\bibliography{reference} 

\newpage 

\begin{center}
    Supplementary Appendix\\
    Comparing the one-sided and monotonic manipulations
\end{center}
 \label{apx.monotone}
    \cite{mccraryManipulationRunningVariable2008} conjectures that ``The density test [...] is expected to be powerful when manipulation is monotonic'' (page. 711); that is, in our notations, $R \geq R^*(0)$ or $R \leq R^*(0)$ almost surely. Although monotonic manipulation is intuitive and shares similarities with our one-sided restriction, they differ. Monotonic manipulation is neither sufficient nor necessary to detect manipulation using the density test. First, monotonic manipulation is not necessary for detection using the density test. Specifically, it imposes unnecessary restrictions on imprecise manipulators that satisfy both \eqref{eq.control} and \eqref{eq.selection}. Unlike monotonic manipulation, imprecise manipulation may follow both incentives to receive and not to receive treatment.
    
    Second, and more importantly, monotonic manipulation is not sufficient to validate the density test. 
    \begin{figure}
    \centering
    \includegraphics[width=0.7\hsize]{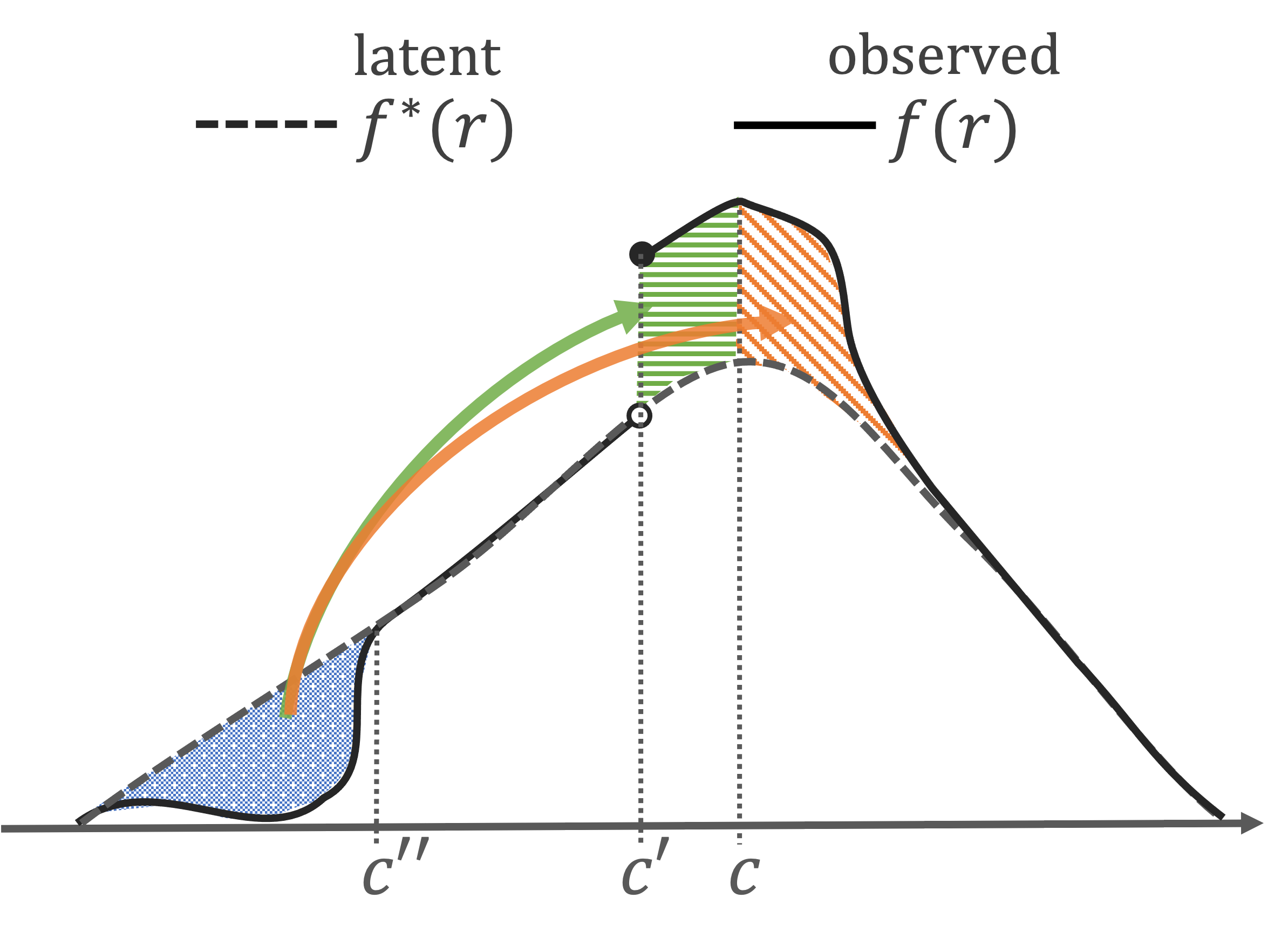}
    \caption*{Counterexample of monotone manipulation violating detectability.}
    \label{fig.nondetectableMonotonicity}        
    \end{figure}
     We provide a counterexample below. Consider teachers offering a retake of an examination. Now consider that some students intentionally failed to ensure that they would retake the examination. In Figure \ref{fig.nondetectableMonotonicity}, there are two nearby cutoffs, $c$ and $c'$. Let the known cutoff $c$ be the official passing cutoff and the left cutoff $c'$ be a hidden cutoff that is unknown to the researcher but is an internal cutoff for penalizing students below it. Some students who were not ready for the examination and had their $R^*(0)$ lower than $c''$ (in the blue-dotted region) may demonstrate two different manipulation behaviors. Some may precisely assign their $R^*(1)$ just above the passing cutoff $c$ (in the hatched region) because they believe they will not improve their scores. Others may intentionally fail the examination to retake it because they think they can score higher than just above the cutoff $c$. Nevertheless, these students set $R^*(1)$ above $c'$ (in the striped region) to avoid facing an additional penalty. These two manipulations are both \textit{monotonic} but not \textit{one-sided} because they are sorted both just below and above the policy cutoff $c$. These monotonic manipulations make the density continuous at $c$. However, students' expected performance in their retaken examination differs from those just above and below the cutoff $c$; consequently, the identification fails.

\end{document}